\documentclass[preprint,prd,tightenlines,nofootinbib]{revtex4}
\usepackage{graphicx}
\font\symbolfont=cmsy10 at 10 pt
\def\texttilde{{\symbolfont \char'030}}
\def \MSbar {\vbox{\hrule\kern 1pt\hbox{\rm MS}}}
\begin{document}

\title{Next-to-leading order QCD calculations with parton showers I:
collinear singularities}
\author{ Michael Kr\"amer}
\affiliation{School of Physics, 
 The University of Edinburgh, Edinburgh EH9 3JZ, Scotland}
\author{ Davison E.\ Soper}
\affiliation{Institute of Theoretical Science, 
University of Oregon, Eugene, OR 97403 USA}
\date{19 February 2004}

\begin{abstract}
Programs that calculate observables in quantum chromodynamics at
next-to-leading order typically suffer from the problem that, when
considered as event generators, the events generated consist of partons
rather than hadrons and just a few partons at that. Thus the events
generated are completely unrealistic. These programs would
be much more useful if the few partons were turned into parton showers,
in the style of standard Monte Carlo event generators. Then the parton
showers could be given to one of the Monte Carlo event generators to
produce hadron showers. We show how to generate parton
showers related to the final state collinear singularities of the
next-to-leading order perturbative calculation for the example of $e^+ +
e^- \to 3\ {\rm jets}$. The soft singularities are left for a separate
publication.
\end{abstract}

\maketitle

\section{Introduction}

Perturbation theory is a powerful tool for deriving the consequences of
the standard model and its extensions in order to compare to experimental
results. When strongly interacting particles are involved, perturbation
theory becomes relevant when the experiment involves a short distance
process, with a momentum scale $Q$ that is large compared to 1 GeV. Then
one can expand the theoretical quantities for the short distance part of
the reaction in powers of $\alpha_s(Q)$, which is small for large $Q$. Of
course, the particles continue to interact at long distances, but for
suitable types of experiments the long distance effects in the final
state can be neglected while the long distance effects in the initial
state can be factored into parton distribution functions that describe the
distributions of partons in incoming hadrons. This works if the
measurement of properties of the final state is {\it infrared safe}: the
measurement is only weakly affected by long distance interactions in the
final state. Examples include the inclusive production of very heavy
particles and the production of (suitably defined) jets of light
particles.

The simplest sort of calculation for this purpose consists of calculating
the hard process cross section at the lowest order in $\alpha_s$, call it
$\alpha_s^B$, at which it occurs. For example, for two-jet production
in hadron-hadron collisions, $B=2$ while for three-jet production in
electron-positron annihilation $B = 1$. This kind of lowest order (LO)
calculation is simple, but leaves out numerically significant
contributions. Corrections from higher order graphs are often found
to be 50\% of the lowest order result. For this reason, one often
performs a next-to-leading order (NLO) calculation, including terms
proportional to $\alpha_s^B$ and $\alpha_s^{B+1}$. Then the estimated
error from yet higher order terms (which are usually unknown) is
typically smaller, say 10\%.

There is another kind of calculational technique available, that of the
Monte Carlo event generator \cite{Pythia, Herwig, Ariadne}. In this
technique, a computer program generates ``events,'' a list of particles
($\pi$, $\rho$, $p$, \dots) and their momenta $p_1,p_2,p_3,\dots$ that
constitute a final state that could arise from the given collision. Call
the $n$th final state $f_{\!n}$. The program is constructed in such a way
that the probability of generating a final state $f$ is approximately
proportional to the physical probability of getting this same final state
in the standard model or in whatever theory is to be tested. That is, the
predicted value for an observable described by a measurement function
${\cal S}(f)$ is
\begin{equation}
\langle {\cal S}\rangle = { 1 \over N}\sum_{n=1}^N
{\cal S}(f_{\!n}).
\end{equation}
Alternatively, the program may generate weights $w_n$ for the events, so
that
\begin{equation}
\langle {\cal S}\rangle = { 1 \over N}\sum_{n=1}^N
w_n\,{\cal S}(f_{\!n}).
\end{equation}
The Monte Carlo event generators have proven to be of enormous value.
Among their benefits is the following.
Since they generate complete final states, one can use detector
simulations to analyze the effects on the measurement of any departure of
the detector from an ideal detector. Essentially, this amounts to
replacing an idealized measurement function ${\cal S}(f)$ by something
that is much closer to what an actual detector does.

There are, of course, some weaknesses in Monte Carlo event generators,
considered as tools to provide predictions of the standard model and its
extensions. One is that the theory encoded in them is more than the
standard model. They also contain models and approximations for
long and medium distance physics. Fortunately, the long and medium
distance effects do not matter much in the case that the measurement
function ${\cal S}(f)$ is infrared safe and involves only one large
momentum scale $Q$.  In this case, the place where the medium and long
distance approximations are important is in assessing the effects of
detector imperfections, which likely result in having an effective
measurement function, ${\cal S}_{\rm eff}(f)$, including the detector
effects, that departs somewhat from the ideal function ${\cal S}(f)$ and
thus departs from being truly infrared safe.

A more serious weakness of current Monte Carlo event generators is that
they are based on leading order perturbation theory, so that a rather
substantial theoretical uncertainty should be ascribed to the result if
the result is considered as deduction from the standard model (or one of
its extensions) with no other hypotheses included.

Return now to the NLO calculations. They are typically computer programs
that act as Monte Carlo generators of partonic events. They produce
events $f$ consisting of a few partons with specified momenta. For
example, for three-jet production in electron-positron annihilation,
there are three or four partons in the final state. Each event comes
with a weight $w_n$. Then the predicted value for an observable described
by a measurement function
${\cal S}(f)$ is
\begin{equation}
\langle {\cal S}\rangle = { 1 \over N}\sum_{n=1}^N
w_n\,{\cal S}(f_{\!n}).
\label{average}
\end{equation}
Here $S(f)$ is an idealized measurement function applied to the partonic
final state. We see that the NLO programs are actually quite close in
structure to the Monte Carlo event generators.

One perceived weakness of the NLO Monte Carlo generators is that the
weights are negative as well as positive. This feature arises because the
programs do real quantum mechanics, in which an amplitude times a complex
conjugate amplitude has a real part that can be either positive or
negative. Negative weights do not, however, cause a problem when inserted
into Eq.~(\ref{average}): computers can easily multiply negative
numbers.

The real weakness of most current NLO Monte Carlo generators is that they
produce simulated final states that are not close to physical final
states. What could be done to generate realistic final states? Following
the example of the leading order event generators, which do generate
realistic final states, one would use the final states in the NLO
partonic generators to generate a parton shower from each outgoing
parton. This gives a final state with lots of partons. Then one could use
one of the highly successful algorithms of the leading order Monte Carlo
event generators to turn the many partons into hadrons. There should not
be a serious problem with the hadronization stage, since this is modeled
as a long distance process that leaves infrared safe observables largely
unchanged. The problem lies with the parton showers. Here a high energy
parton splits into two daughter partons, which each split into two more
partons. As this process continues, the virtualities of successive pairs
of daughter partons gets smaller and smaller, representing splittings
that happen at larger and larger distance scales. The late splittings
leave infrared safe observables largely unchanged. However, the first
splittings in a parton shower can involve large virtualities. They
represent a mixture of long distance and short distance physics. Thus one
must be careful that the showering does not reproduce some piece of short
distance physics that was already included in the NLO calculation. 

Consider an example that illustrates these ideas, jet production in $pp$
collisions with $\sqrt s = 14\ {\rm TeV}$ at the Large Hadron Collider
now under construction. One can calculate the inclusive cross section
$d\sigma/dM_{JJ}$ to make two jets with a dijet mass $M_{JJ}$. We may
specify that the jets are defined with the $k_T$ algorithm
\cite{kTjets}, that we integrate over the rapidities of the two jets
in the range $|y_1|<1, |y_2|<1$, and that $M_{JJ}$ is in the range $0.5\
{\rm TeV} < M_{JJ} < 5\ {\rm TeV}$. There are computer programs currently
available to perform this calculation at NLO \cite{EKS,giele,nagy}. There
are no large logarithms in the theory in this kinematic region. Thus an
NLO calculation should be reliable. If the experiment were to disagree
with the theory beyond the experimental errors and the estimated errors
from uncalculated NNLO contributions and from parton distributions, that
would be a significant indication that some sort of physics not
included in the standard model is at work. 

Although the theoretical picture just outlined seems bright, the
available programs do not produce realistic final states.  To illustrate
this, imagine computing a cross section that is sensitive to the final
state. For example, one might use a standard NLO program to calculate the
cross section $d\sigma/dM_{JJ}dM_1$, where $M_1$ is the mass of the jet
with the larger rapidity. Our result would {\em not} be sensible in the
region $M_1 \ll M_{JJ}$. It would have a nonintegrable singularity at
$M_1 = 0$ together with a term proportional to $\delta(M_1)$ with an
infinite negative coefficient. The integral over $M_1$ of this highly
singular function would give back $d\sigma/dM_{JJ}$, but the differential
distribution would be highly nonphysical. In contrast, the standard Monte
Carlo event generators can generate sensible results simultaneously for
both $d\sigma/dM_{JJ}$ and $d\sigma/dM_{JJ}\,dM_1$. The only drawback is
that the result for $d\sigma/dM_{JJ}$ will be correct only to leading
order in an expansion in powers of $\alpha_s(M_{JJ})$. The problem
addressed in this paper (using, however, an example from
electron-positron annihilation) is how to modify such an NLO calculation
so as to take advantage of the parton shower technology built into Monte
Carlo event generators. We wish to get quantities like
$d\sigma/dM_{JJ}\,dM_1$ approximately right while not losing the NLO
accuracy of $d\sigma/dM_{JJ} = \int dM_1\ d\sigma /dM_{JJ}\,dM_1$.

Some technical language is helpful here in order to make these ideas more
precise. Using the familiar concept of infrared safety (defined precisely
in Sec.~\ref{sec:NLO}), one can describe the cross section
$d\sigma/dM_{JJ}$ as an infrared safe two-jet cross section. Then the
cross section $d\sigma/dM_{JJ}dM_1$ for $M_1 > 0$ is an infrared safe
three-jet cross section: it is infrared safe and is nonzero only when the
number of final state particles is three or more. With this language,
what we wish to do is get two-jet cross sections right to NLO and at the
same time get $n$-jet cross sections approximately right for $n>2$.

Recall that the shower Monte Carlo programs work by using an
approximate version of the perturbative theory at an infinite number of
orders in perturbation theory, approximating the behavior of the Feynman
graphs near the singularities corresponding to collinear parton splitting
or soft gluon emission. These singularities lead to large logarithms, two
powers of $\log(M_1/M_{JJ})$ for each additional power of $\alpha_s$. The
programs approximately add up a series of $\alpha_s^N$ factors times
logarithmic factors. After this, there is a phase that models how partons
turn into hadrons. In this paper we omit the hadronization phase and use
our own code for the parton showering. Our intention is that in the
future only the stages of showering that directly connect to the hard
scattering would be part of a program for NLO hard scattering with
showers. The parts of the shower related to ``softer'' parton splittings,
along with hadronization, would be performed by one of the standard Monte
Carlo programs. We thus concentrate in this paper on the
showering--hard-scattering interface and not on the showering itself.

We can illustrate the same points for an NLO base calculation that has one
more jet.  An instructive example is the cross section discussed above,
$d\sigma/dM_{JJ}\,dM_1$. Here we do not integrate over
$M_1$ and we insist that $M_1/M_{JJ}$ not be small. This is technically
an infrared safe three-jet cross section, as discussed earlier. In a
three or four parton system that appears in the calculation of this
quantity at NLO, jet 1 consists of two or three partons and has mass
$M_1$, while jet 2 consists of one or two partons. The NLO calculation
(using \cite{nagy}) should be reliable but would not have sensible final
states. Thus if we let $M_2$ be the mass of whichever of the two measured
jets has the {\it smaller} rapidity and tried to calculate
$d\sigma/dM_{JJ}\,dM_1\,dM_2$, the result would not be even approximately
correct for $M_2 \ll M_{JJ}$ and $M_2 \ll M_1$. One would then like to
have a method that could get quantities like $d\sigma/dM_{JJ}\,dM_1\,dM_2$
approximately right while not loosing the NLO accuracy of
$d\sigma/dM_{JJ}\,dM_1$. 

This discussion can serve to illustrate another point.  Note that a
calculation of $d\sigma/dM_{JJ}\,dM_1$ will have a problem if
$M_{1}/M_{JJ} \ll 1$ because the result will contain large logarithms,
namely logs of $M_{1}/M_{JJ}$. The large logarithms are associated with
the partons getting close to a two-jet configuration. One could then view
the event as arising from a two-jet event in which one of the jets splits
into two nearly collinear jets or one hard jet and one soft jet.  A
two-jet calculation with showers would get the effects of these large
logarithms approximately right. What one would then want is a method to
merge the two-jet and three-jet calculations so that 1) a two-jet
infrared safe observable is correctly calculated at order $\alpha_s^3$,
{i.e.}\ NLO, 2) a three-jet infrared safe observable with suitable
cuts so that the contributing events are well away from the two-jet
region is correctly calculated at order $\alpha_s^4$, {i.e.}\ NLO for
three-jets, and 3) a three-jet infrared safe observable with cuts that
allow contributions from close to the two-jet region is correctly
calculated at order $\alpha_s^4$ with additional corrections proportional
to $\alpha_s^{n+4}$ times logs, with $n \ge 1$, that approximately
account for the production of three jets by showering from a two-jet
event. An investigation of this merging problem would certainly be
desirable, but is beyond the scope of this paper. 

This paper concerns $e^+ + e^- \to 3\ {\rm {jets}}$. The goal is
to see how to modify a NLO calculation by adding parton
showers in such a way that when the generated events are used to
calculate an infrared safe observable, the observable is calculated
correctly at next-to-leading order. To do this, we build on our work
in \cite{KSCoulomb}, which sets up the NLO calculation in the Coulomb
gauge. It is convenient to use a physical gauge for this purpose because
it makes the analysis simpler. In this paper we explain the aspects of
this problem that relate to the divergences that appear in perturbation
theory when two massless partons become collinear. This part of the
method is, we think, quite simple and easy to understand. The treatment
of the divergences that appear when a gluon becomes soft is a bit more
technical and will be given in a separate paper
\cite{softshower}.

The algorithms that we propose apply to showers from final state massless
partons. They are illustrated using the process $e^+ + e^- \to 3\ {\rm
{jets}}$ in quantum chromodynamics and are embodied in computer code that
performs calculations for this process \cite{beowulfcode}. We do not
discuss showers from initial state massless partons. While our research
program (starting with \cite{KSCoulomb}) was underway, Frixione and
Webber succeeded in matching showers to a NLO calculation for a problem
with massless partons in the initial state (but no observed colored
partons in the final state) \cite{FrixioneWebberI}. Subsequently,
Frixione, Webber, and Nason have extended this method to massive colored
partons in the final state \cite{FrixioneWebberII}. The general approach
of
\cite{FrixioneWebberI,FrixioneWebberII} is quite similar to that of the
present paper with respect to collinear singularities. The principle
difference is that we generate the first splittings of partons from an
order $\alpha_s^B$ diagram using the functions given by the partonic
self-energy diagrams in the Coulomb gauge, while \cite{FrixioneWebberI,
FrixioneWebberII} uses the splitting functions of the Herwig Monte Carlo
event generator.

We discuss $e^+ + e^- \to 3\ {\rm {jets}}$ but not $e^+ + e^- \to 2\
{\rm {jets}}$. Of course, to calculate $e^+ + e^- \to 2\ {\rm {jets}}$ at
order $\alpha_s$ with parton showering added would be quite a lot simpler
than the present calculation for $e^+ + e^- \to 3\ {\rm {jets}}$ at
order $\alpha_s^2$. However, merging the two calculations would not be
trivial and is not attempted here. We simply suppose that the present
calculation is to be applied with a measurement function that gives zero
for events that are close to the two-jet configuration (for example, for
thrust near one).

There has been other recent work on this problem \cite{otherwork}. That of
Collins \cite{JCC} is particularly instructive. This approach seeks both
to put the parton showering calculation on a more rigorous basis and at
the same time to match it to the hard scattering calculation, viewing the
hard scattering as one end of a cascade of virtualities. Our goal is more
limited. We take it as proven by experience that parton showers work well
to make reasonably realistic final states. Our only concern is to make
sure that when we add parton showers we do not lose the NLO accuracy of
the result.

\section{Next-to-leading order calculation}
\label{sec:NLO}

We begin with a program that calculates an infrared safe observable in
electron-positron annihilation at next-to-leading order. The program
\cite{KSCoulomb} uses a physical gauge, specifically the Coulomb gauge,
and performs all of the integrals, including the virtual loop integrals,
numerically \cite{beowulfPRL, beowulfPRD, beowulfrho}. It is not really
necessary to perform the virtual loop integrals numerically. It is a lot
easier, but if one wanted to perform these integrals by hand ahead of
time and insert the analytical results into the program, the calculation
could still proceed. It is also not necessary to use a physical gauge.
However, the use of a physical gauge makes the method conceptually
simpler. In particular, collinear singularities are isolated in very
simple Feynman subdiagrams, the cut and uncut two-point functions. If one
were to use the Feynman gauge, the collinear singularities would be spread
over many kinds of graphs and one would need another layer of analysis to
organize them into a simple form.
The Coulomb gauge, in particular, is well suited for calculations of
electron-positron annihilation, in which the physics is approximately
rotationally invariant in the electron-positron c.m.\ frame. The
time-like axial gauge has this same good property, but is not well suited
for our algorithm for a technical reason: it introduces singularities in
the complex energy plane for loop integrals, which would make the
evaluation of the energy integrals in our algorithm more complicated.

In order to present a method for adding showers to the NLO program, we
need to review a little about how the NLO calculation works. We present
the barest outline. Details may be found in \cite{beowulfPRL, beowulfPRD,
beowulfrho, KSCoulomb}. In the NLO program, the observable is expressed
in the form
\begin{eqnarray}
{\cal I} &=& 
\sum_n { 1 \over n!}\,\int d\vec p_1\cdots d\vec p_n\
\delta\left(\sum\vec p_i\right)\,
{ h\!\left(\sqrt{s}\right) \over \sigma_0\left(\sqrt s\, \right)}\,
\nonumber\\
&&\times 
F_{\!n}\!\left(\vec p_1,\dots,\vec p_n,\sqrt s,
\alpha_s\!\left(\sqrt S\right)\right) 
{\cal S}_n\!\left(\vec p_1\sqrt{S/s},\dots ,\vec p_n\sqrt{S/s}\right).
\label{integrationstruct}
\end{eqnarray}
Here there is a sum over the number $n$ of final state partons. For a
next-to-leading order calculation of a three-jet observable, $n$  runs
from 3 to 4. The c.m.\ energy is $\sqrt S$. There are integrations over
dimensionless momenta $\vec p_j$ for the final state partons. The true,
dimensionful  momenta are $\vec P_j = \vec p_j\sqrt{S/s}$ where $\sqrt
s = \sum |\vec p_j|$. There is a delta function for momentum
conservation. In place of the delta function for energy conservation,
there is a function $h(\sqrt s)$ that satisfies $\int d \sqrt s\
h(\sqrt s) = 1$. There is a conventional normalizing  factor ${ 1 /
\sigma_0\left(\sqrt s\, \right)}$, where $\sigma_0$ is the Born cross
section for $e^+ + e^- \to {\rm hadrons}$. 

\begin{figure}[ht]
\includegraphics[width = 8 cm]{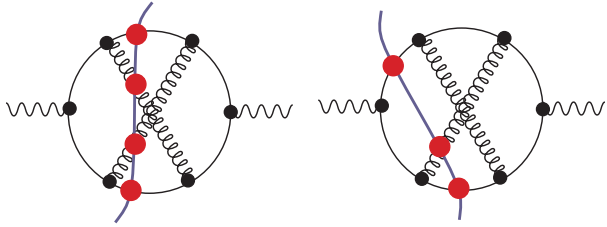}
\medskip
\caption{Two cuts of one of the Feynman diagrams that contribute to 
$e^+e^- \to {\rm hadrons}$.}
\label{fig:cutdiagrams}
\end{figure}

The functions $F_{\!n}$ contain the perturbative matrix elements,
calculated from the Feynman diagrams for the process. For instance, two
such diagrams are depicted in Fig.~\ref{fig:cutdiagrams}. The first has
four partons in the final state and contributes to $F_{\!4}$, while the
second has three partons in the final state and contributes to
$F_{\!3}$. In the case of diagrams with a virtual loop, as in the second
diagram in Fig.~\ref{fig:cutdiagrams}, the function $F_{\!n}$ contains an
integral over a dimensionless loop momentum $\vec l$. It is an essential
part of the numerical method used here that the integration over the loop
momentum is performed numerically, at the same time as the integrations
over the final state momenta are performed.

At next-to-leading order, the functions $F_{\!n}$ contain either one or
two powers of $\alpha_s$. We take the renormalization scale to be $\mu
= c_\mu \sqrt S$, where $c_\mu$ is a dimensionless parameter. In our
numerical example at the end of this paper, we take $c_\mu = 1/6$.

The last ingredients in Eq.~(\ref{integrationstruct}) are the
measurement functions ${\cal S}_n$, which are functions of the 
dimensionful final state momenta $\vec p_j\sqrt{S/s}$. These functions
are depicted in Fig.~\ref{fig:cutdiagrams} by dots where parton
lines cross the cut that represents the final state. The functions
${\cal S}_n$ represent a three-jet observable in the sense that ${\cal
S}_2 = 0$.  They are invariant under interchanges of their arguments
and are infrared safe in the sense that
\begin{equation}
{\cal S}_{n+1}\!\left(\vec P_1,\dots, \lambda\vec P_n,
(1-\lambda)\vec P_n\right) 
= {\cal S}_{n}\!\left(\vec P_1,\dots,\vec P_n\right)
\label{irsafe}
\end{equation}
for $0 \le \lambda < 1$, so that the measurement is unchanged if a
parton splits into two partons moving in the same direction.

\section{One loop self-energy diagrams}
\label{sec:selfenergy}

In order to formulate parton showering properly, we need to understand
how one parton splits into two at the one loop level in the Coulomb gauge.
In this section, we analyze the perturbative calculation of a cut
quark propagator that occurs as part of a larger graph. The analysis for
gluons is essentially the same.

To set the notation, consider first the contribution at order
$\alpha_s^0$:
\begin{equation}
{\cal I}[{\rm Born}] = \int\! {d\vec q\over  2|\vec q\,|}\ {\rm
Tr}\left\{ \rlap{/} q \, R_0 
\right\},
\label{quarkborn}
\end{equation}
where $\vec q$ is the three-momentum carried by the propagator, $q^\mu =
(|\vec q\,|,\vec q\,)$,  and $R_0$ denotes the factors associated with the
rest of the graph and with the final state measurement function. In
particular, $R_0$ has two Dirac indices that match the Dirac indices on
the $\rlap{/} q$ and there is a trace over these indices.

Next, consider the contribution from a cut self-energy insertion on the
quark propagator in question. The quark splits into a quark plus a
gluon, giving a contribution of the form
\begin{equation}
{\cal I}[{\rm real}] =
\int\! {d\vec q\over  2|\vec q\,|}\ {\rm Tr}\left\{ 
\int_0^\infty\! {d\bar q^2 \over \bar q^2}
\int_0^1\!dx
\int_{-\pi}^\pi\! {d\phi\over 2\pi}\
{\alpha_s\over 2\pi}{\cal M}_{g/q}(\bar q^2,x,\phi)\,
R(\bar q^2,x,\phi) 
\right\}.
\label{quarkreal}
\end{equation}

Here there is an integration over the loop momentum that
describes the splitting. We use coordinates $\left\{\bar q^2, x,
\phi\right\}$ for this loop momentum. We define these as
follows. Let the gluon momentum be
$\vec k_+$ and the quark momentum after the splitting be $\vec k_-$,
with $\vec k_+ + \vec k_- = \vec q$. We define $\bar q^2$ and $x$ by
\begin{eqnarray}
\sqrt{1 + \bar q^2/|\vec q\,|^2} &=& 
(|\vec k_+| + |\vec k_-|)/|\vec q\,| ,
\nonumber\\
2 x - 1 &=& (|\vec k_+| - |\vec k_-|)/|\vec q\,|.
\end{eqnarray}
Then $0 < \bar q^2$ and $0<x<1$. The variables $\bar q^2$ and $x$ are
constant on surfaces that are, respectively, ellipsoids and
hyperboloids in loop momentum space. These surfaces are orthogonal to
each other where they intersect. The virtuality of the quark-gluon pair
is
\begin{equation}
(|\vec k_+| + |\vec k_-|)^2 - |\vec q\,|^2
= \bar q^2.
\end{equation}
Finally, we let $\phi$ be the azimuthal angle of $\vec k_+$ in a
coordinate frame in which the $z$ axis lies along $\vec q$ and the $x$
axis is defined arbitrarily.

The function $[{\alpha_s/(2\pi)}]{\cal M}_{g/q}(\bar q^2,x,\phi)$ in
Eq.~(\ref{quarkreal}) is determined directly from the Feynman rules and
is given as ${\cal M}_{q}$ in Eq.~(87) of
Ref.~\cite{KSCoulomb}.\footnote{For the purposes of this paper, we thought
it best to display the factor $\alpha_s$ explicitly.} Its $\bar q^2 \to
0$ limit is simple:
\begin{equation}
{\alpha_s\over 2\pi}{\cal M}_{g/q}(\bar q^2,x,\phi) \sim
\rlap{/}q\,{ \alpha_s \over 2\pi}\,
\tilde P_{g/q}(x),
\label{calMqtoAP}
\end{equation}
where $\tilde P_{g/q}(x)$ is the one loop Altarelli-Parisi kernel for
$q \to g$, namely $C_F[1 + (1-x)^2]/x$. [We follow the notation of
Ref.~4, in which $\tilde P_{a/b}(x)$ denotes the Altarelli-Parisi kernel
for $b \to a$ but omitting regulation of the $x \to 1$ singularity, if
there is such a singularity for that $\{a,b\}$ combination.]

The function $R$ denotes the rest of the graph and the measurement
function, as before. Because it includes the measurement function, it
depends on $x$ and $\phi$ as well as on $\bar q^2$. Because the
measurement function is infrared safe, the $\bar q^2\to 0$ limit of $R$
is simple:
\begin{equation}
R(\bar q^2,x,\phi) \to R_0
\hskip 1 cm {\rm as}\hskip 1 cm \bar q^2 \to 0,
\label{Rlimit}
\end{equation}
where $R_0$ is the function appearing in the Born cross section,
Eq.~(\ref{quarkborn}).

Next, consider the contribution from virtual self-energy insertions on
the quark propagator in question. The quark splits into a quark plus a
gluon that recombine, leaving a single quark line in the final state.
The net contribution from the two virtual loop graphs is
\begin{equation}
{\cal I}[{\rm virtual}] =
\int\! {d\vec q\over  2|\vec q\,|}\ {\rm Tr}\left\{
- 
\int_0^\infty {d\bar q^2 \over \bar q^2}
\int_0^1\! dx
\int_{-\pi}^\pi\! {d\phi\over 2\pi}\ 
{ \alpha_s \over 2\pi}\,
{\cal P}_{\!g/q}(\bar q^2,x)\,\rlap{/}q\, R_0 
\right\}.
\label{quarkvirt}
\end{equation}
Here, again, $q^\mu = (|\vec q\,|,\vec q\,)$ is the on-shell incoming
momentum. The function ${\cal P}_{\!g/q}$ is given in
the Appendix. It equals the function ${\cal P}_{\!q}$ calculated
in Ref.~\cite{KSCoulomb} except for modifications to the terms that
reflect renormalization. The most crucial feature of ${\cal
P}_{\!g/q}$ is its $\bar q^2 \to 0$ limit:
\begin{equation}
{\cal P}_{\!g/q}(\bar q^2,x) \to
\tilde P_{g/q}(x)
\hskip 1 cm {\rm as}\hskip 1 cm \bar q^2 \to 0.
\label{calPqtoAP}
\end{equation}

\begin{figure}
\centerline{
\includegraphics[width = 15 cm]{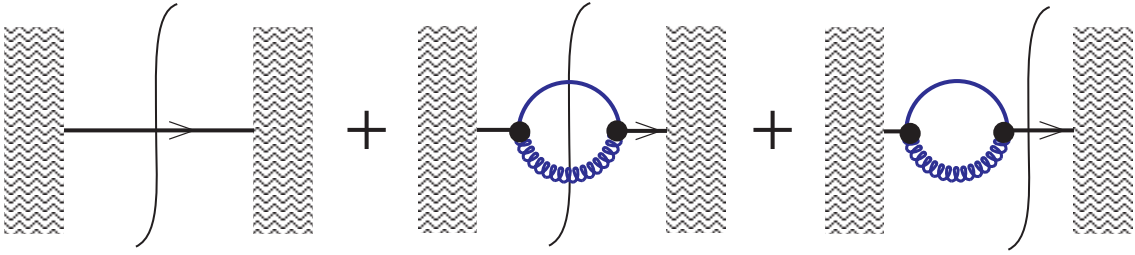}\ \ \
}
\medskip
\caption{Graphical representation of the three terms in Eq.~(\ref{pert}).
The third diagram is intended to represent the entire virtual
contribution.}
\label{fig:threeterms}
\end{figure}

If we combine the three perturbative contributions, we have
\begin{eqnarray}
{\cal I}[{\rm Born}]+
{\cal I}[{\rm real}] + {\cal I}[{\rm virtual}]
&=&\int\! {d\vec q\over  2|\vec q\,|}\ {\rm Tr}\biggl\{
\rlap{/} q \, R_0 
+
\int_0^\infty {d\bar q^2 \over \bar q^2}
\int_0^1\! dx
\int_{-\pi}^\pi\! {d\phi\over 2\pi}\
\nonumber\\
&&
\left[
{\alpha_s\over 2\pi}{\cal M}_{g/q}(\bar q^2,x,\phi)
R(\bar q^2,x,\phi)
-
{ \alpha_s \over 2\pi}\,{\cal P}_{\!g/q}(\bar q^2,x)\,\rlap{/}q\,
R_0
\right]
\biggr\}.\ \ \ \ 
\label{pert}
\end{eqnarray}
These three terms are represented in Fig.~\ref{fig:threeterms}.
The real and virtual contributions are separately infrared divergent.
However the integrands are added before the integration is performed
numerically. Then the divergence cancels because of
Eqs.~(\ref{calMqtoAP}), (\ref{calPqtoAP}), and (\ref{Rlimit}). (It is
also important for the cancellation that the two terms match when $\bar
q^2 \to 0$ and $x \to 0$ at the same time, with $\bar q^2/x$ fixed.)

\section{One level shower for a single parton}
\label{sec:onelevel}

\begin{figure}
\centerline{
\includegraphics[width = 4.06 cm]{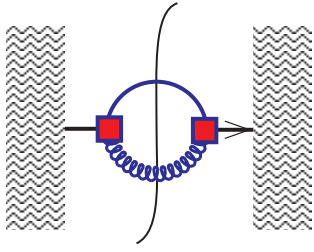}
}
\medskip
\caption{Graphical representation of ${\cal I}[{\rm shower}]$ in
Eq.~\ref{shower}. The pair of square vertices represents the matrix
element ${\cal M}$ and the Sudakov exponential.}
\label{fig:showersplitting}
\end{figure}

So far, we see how a next-to-leading order numerical calculation works,
but there is no parton shower. If we wanted to have the first splitting
in a parton shower, we could follow the spirit of the parton shower
Monte Carlo programs \cite{Pythia, Herwig, Ariadne} and replace
Eq.~(\ref{pert}) by
\begin{eqnarray}
{\cal I}[{\rm shower}]&=&
\int\! {d\vec q\over  2|\vec q\,|}\ {\rm Tr}\Biggl\{ 
\int_0^\infty {d\bar q^2 \over \bar q^2}
\int_0^1\! dx
\int_{-\pi}^\pi\! {d\phi\over 2\pi}\
{\alpha_s\over 2\pi}{\cal M}_{g/q}(\bar q^2,x,\phi)
R(\bar q^2,x,\phi)
\nonumber\\
&&\times
\exp\left(-\int_{\bar q^2}^\infty{ d\bar l^2 \over \bar l^2}
\int_0^1\! dz\
{ \alpha_s \over 2\pi}\,{\cal P}_{\!g/q}(\bar l^2, z)
\right)
\Biggr\}.
\label{shower}
\end{eqnarray}
The quantity ${\cal I}[{\rm shower}]$ is represented graphically in
Fig.~\ref{fig:showersplitting}.

In Eq.~(\ref{shower}), we keep the construction simple by maintaining
$\alpha_s$ evaluated at a fixed scale $\mu$. We choose not to take the
$\bar q^2 \to 0$ limit in any part of the function $R(\bar q^2,x,\phi)$
that describes the rest of the graph. The parton splits with the exact
matrix element ${\cal M}_{g/q}$, whereas a typical Monte Carlo program
would use the small $\bar q^2$ limit of ${\cal M}_{g/q}$. As is standard,
the exponential factor is interpreted as the probability for the parton
not to split at virtuality greater than $\bar q^2$. In place of ${\cal
P}_{\!g/q}$ in the exponent, one might use the Altarelli-Parisi
kernel supplemented by some theta functions that keep $x$ away from 0 and
1, but we use the exact function ${\cal P}_{\!g/q}$ that occurs in the
virtual corrections.

We see here that parton shower calculations as exemplified in
Eq.~(\ref{shower}) are very different from fixed order calculations as
exemplified in Eq.~(\ref{pert}). In the fixed order calculation, the
virtual graph acts as a subtraction to the real splitting contribution.
In the shower calculation, the virtual graph becomes a multiplicative
modification to the splitting matrix element. 

The perturbative representation in Eq.~(\ref{pert}) is
the same as the shower representation in Eq.~(\ref{shower}) up to
corrections of order $\alpha_s^2$:
\begin{equation}
{\cal I}[{\rm Born}]+
{\cal I}[{\rm real}]+
{\cal I}[{\rm virtual}]
= {\cal I}[{\rm shower}]\times\left(1 + {\cal O}(\alpha_s^2)\right).
\label{theorem}
\end{equation}
The general idea incorporated in this {\it subtraction to multiplication
theorem} is a standard part of the strategy behind shower Monte Carlo
programs,\footnote{See, for example, Eq.~(B.15) in Appendix B of
Ref.~\cite{FrixioneWebberI}.} but we have not found the theorem stated in
just this way, so we prove it here. First, we write
${\cal I}[{\rm shower}]$ as
\begin{equation}
{\cal I}[{\rm shower}] = 
{\cal I}_1[{\rm shower}] + {\cal I}_2[{\rm shower}]
\end{equation}
with
\begin{eqnarray}
{\cal I}_1[{\rm shower}] &=&
\int\! {d\vec q\over  2|\vec q\,|}\ {\rm Tr}\Biggl\{
\int_0^\infty {d\bar q^2 \over \bar q^2}
\int_0^1\! dx
\int_{-\pi}^\pi\! {d\phi\over 2\pi}\
\exp\left(-\int_{\bar q^2}^\infty
{ d\bar l^2 \over \bar l^2}\int_0^1\! dz\
{ \alpha_s \over 2\pi}\,{\cal P}_{\!g/q}(\bar l^2,z)\right)
\nonumber\\
&&\times
\biggl[
{\alpha_s\over 2\pi}{\cal M}_{g/q}(\bar q^2,x,\phi)\,R(\bar q^2,x,\phi)
-
{ \alpha_s \over 2\pi}\,{\cal P}_{\!g/q}(\bar q^2,x)\,\rlap{/}
q\,
R_0\biggr]
\Biggr\}
\label{Ishower1}
\end{eqnarray}
and
\begin{eqnarray}
{\cal I}_2[{\rm shower}] &=&
\int\! {d\vec q\over  2|\vec q\,|}\ {\rm Tr}\Biggl\{
\rlap{/}q\,R_0
\int_0^\infty \! d\bar q^2 \
\nonumber\\
&&\hskip 1 cm\times
{ 1 \over \bar q^2}
\int_0^1\! dx\
{ \alpha_s \over 2\pi}\,{\cal P}_{\!g/q}(\bar q^2,x)\
\exp\left(-\int_{\bar q^2}^\infty
{ d\bar l^2 \over \bar l^2}\int_0^1\! dz\
{ \alpha_s \over 2\pi}\,{\cal P}_{\!g/q}(\bar l^2,z)\right)
\Biggr\}.\ \ \ 
\label{Ishower2}
\end{eqnarray}

Let us find the lowest order (up to and including order $\alpha_s\times
R$) terms in the expansions of these integrals. The integral ${\cal
I}_1[{\rm shower}]$ is simple. We take the order $\alpha_s^0$ term in the
exponential to obtain
\begin{eqnarray}
{\cal I}_1[{\rm shower}] &=&
\int\! {d\vec q\over  2|\vec q\,|}\ {\rm Tr}\Biggl\{
\int_0^\infty {d\bar q^2 \over \bar q^2}
\int_0^1\! dx
\int_{-\pi}^\pi\! {d\phi\over 2\pi}\
\nonumber\\
&&\times
\biggl[
{\alpha_s\over 2\pi}{\cal M}_{g/q}(\bar q^2,x,\phi)\,R(\bar q^2,x,\phi)
-
{ \alpha_s \over 2\pi}\,{\cal P}_{\!g/q}(\bar q^2,x)\,\rlap{/}
q\,
R_0\biggr]
\Biggr\}
+{\cal O}(\alpha_s^2\times R).\ \ \
\label{shower1F}
\end{eqnarray}
The integral ${\cal I}_2[{\rm shower}]$ is more subtle. We
cannot simply expand under the integral since the integrals of each term
would be divergent at small $\bar q^2$. Instead, we note that we have
the integral over $\bar q^2$ of a total derivative, so that the
integral is just the difference of the integrand between the
integration end points:
\begin{eqnarray}
{\cal I}_2[{\rm shower}] &=&
\int\! {d\vec q\over  2|\vec q\,|}\ {\rm Tr}\left\{ 
\rlap{/} q\,
R_0
\right\}
[F(\infty) - F(0)],
\label{shower2F0}
\end{eqnarray}
where
\begin{equation}
F(\bar q^2) = \exp\left(-\int_{\bar q^2}^\infty
{ d\bar l^2 \over \bar l^2}\int_0^1\! dz\
{ \alpha_s \over 2\pi}\,{\cal P}_{\!g/q}(\bar l^{\,2},z)\right).
\end{equation}
Now the function ${\cal P}_{\!g/q}(\bar l^{\,2},z)$, which is given in the
Appendix, has the important property that it is proportional to $1/\bar
l^{\,2}$ for $\bar l^{\,2} \to \infty$. This insures that the integral in
the exponent is convergent for $\bar l^{\,2} \to \infty$, so that $F(\bar
q^2) \to \exp(-0) = 1$ as $\bar q^2 \to \infty$. On the other hand, ${\cal
P}_{\!g/q}(\bar l^{\,2},z)$ approaches ${P}_{\!g/q}(z)$ as $\bar l^{\,2}
\to 0$. This makes the $\bar l^{\,2}$-integral diverge as $\bar q^2 \to
0$. In fact, one gets two powers of $\log(\bar q^2)$ for $\bar q^2 \to
0$, one directly from the $\bar l^{\,2}$-integral and one from a
divergence at $z
\to 0$ in the $z$-integral that develops as ${\cal P}_{\!g/q}(\bar l^2,z)$
gets closer and closer to ${P}_{\!g/q}(z)$. The net effect is that the
exponent behaves like a positive constant times $\log^2(\bar q^2)$ for
$\bar q^2 \to 0$. Thus $F(\bar q^2) \to \exp(- \infty) = 0$ as $\bar q^2
\to 0$. The result for ${\cal I}_2[{\rm shower}]$ is then
\begin{eqnarray}
{\cal I}_2[{\rm shower}] &=&
\int\! {d\vec q\over  2|\vec q\,|}\ {\rm Tr}\left\{ 
\rlap{/} q\,
R_0
\right\}.
\label{shower2F}
\end{eqnarray}
Adding the results (\ref{shower1F}) and (\ref{shower2F}), we have the
result in Eq.~(\ref{pert}) plus ${\cal O}(\alpha_s^2\times R)$. This
proves the theorem.

There is an analogous result for the gluon propagator. The trace is
over Lorentz indices instead of Dirac indices. The functions ${\cal
M}_{g/q}$ and ${\cal P}_{\!g/q}$ are replaced by functions 
${\cal M}_{g/g} + N_F {\cal M}_{q/g}$ given\footnote{To be precise,
$[\alpha_s/(2\pi)]({\cal M}_{g/g} + N_F {\cal M}_{q/g})$ is given in
Eq.~(37) of  Ref.~\cite{KSCoulomb} as ${\cal M}_{g}$.} in
Ref.~\cite{KSCoulomb} and ${\cal P}_{\!g/g} + N_F {\cal P}_{\!q/g}$ given
in the Appendix.
 
It should be evident that one can use the subtraction to multiplication
theorem to add parton showers to the next-to-leading order calculation of
three-jet cross sections, while keeping unchanged the first two
terms in the perturbative expansion of the result. Indeed, there is
more than one way to do this. Which way is best can and should be
debated, but in this paper we confine ourselves to mapping out one way.

We will refer to the first splittings of the partons emerging from a Born
graph as primary splittings. The basic idea for the primary splittings is
this.  Start with a computer program that does calculations at
next-to-leading order in the Coulomb gauge. In each cut $\alpha_s^1$
graph, replace ${\cal I}[{\rm Born}]$ by ${\cal I}[{\rm shower}]$ for
each cut propagator. Then take the cut order $\alpha_s^2$ graphs that
have a cut self-energy subgraph plus its corresponding virtual
correction. In each such graph, replace ${\cal I}[{\rm real}] + {\cal
I}[{\rm virtual}]$ by $0$. (There are some complications that arise when
applying this basic idea. We address the complications in subsequent
sections.)

Once one understands this basic idea, one can envisage a lot of
refinements. Here we will pursue one possible refinement. In 
${\cal I}[{\rm shower}]$ we choose a number $\lambda_V$ of order 1 and
divide the integration region into a low virtuality part, $0 < \bar q^2 <
\lambda_V\vec q^{\,2}$, and a high virtuality part $\lambda_V \vec q^{\,2}
< \bar q^2 < \infty$. In the high virtuality term, we expand the Sudakov
exponential in powers of $\alpha_s$ and keep only the order 0 term. This
gives
\begin{equation}
{\cal I}[{\rm shower}] = 
\biggl(
{\cal I}[{\rm shower},\lambda_V]
+ {\cal I}[{\rm real},\lambda_V]
\biggr)
\times\left(1 + {\cal O}(\alpha_s^2)\right)
\label{lambdamodification}
\end{equation}
where
\begin{eqnarray}
{\cal I}[{\rm shower},\lambda_V]&=&
\int\! {d\vec q\over  2|\vec q\,|}\ {\rm Tr}\Biggl\{ 
\int_0^{\lambda_V\vec q^{\,2}}\! {d\bar q^2 \over \bar q^2}
\int_0^1\! dx
\int_{-\pi}^\pi\! {d\phi\over 2\pi}\
{\alpha_s\over 2\pi}{\cal M}_{g/q}(\bar q^2,x,\phi)
R(\bar q^2,x,\phi)
\nonumber\\
&&\times
\exp\left(-\int_{\bar q^2}^\infty
{ d\bar l^2 \over \bar l^2}
\int_0^1\! dz\
{ \alpha_s \over 2\pi}\,{\cal P}_{\!g/q}(\bar l^2, z)
\right)
\Biggr\}.
\label{showerlambda}
\end{eqnarray}
and
\begin{equation}
{\cal I}[{\rm real},\lambda_V] =
\int\! {d\vec q\over  2|\vec q\,|}\ {\rm Tr}\left\{ 
\int_{\lambda_V\vec q^{\,2}}^\infty\! {d\bar q^2 \over \bar q^2}
\int_0^1\!dx
\int_{-\pi}^\pi\! {d\phi\over 2\pi}\
{\alpha_s\over 2\pi}{\cal M}_{g/q}(\bar q^2,x,\phi)\,
R(\bar q^2,x,\phi) 
\right\}.
\label{quarkreallambda}
\end{equation}

Then the refinement of the basic idea for the primary splittings is as
follows.  Start with a computer program that does calculations at
next-to-leading order in the Coulomb gauge. In each cut
$\alpha_s^1$ graph, replace ${\cal I}[{\rm Born}]$ by ${\cal I}[{\rm
shower},\lambda_V]$ for each cut propagator. Then take the cut order
$\alpha_s^2$ graphs that have a cut self-energy subgraph plus its
corresponding virtual correction. In each such graph, replace ${\cal
I}[{\rm real}] + {\cal I}[{\rm virtual}]$ by ${\cal I}[{\rm
real},\lambda_V]$. This approach is a little more complicated than the
simple approach we explained first, which amounts to choosing $\lambda_V
= \infty$.  

Why would one choose $\lambda_V < \infty$? Unless one has a NNLO
calculation to compare to, one cannot know for sure which approach is
best, since the approaches differ only at order $\alpha_s^{B+2}$. However,
the $\lambda_V < \infty$ approach has the advantage that we apply the
formula with Sudakov suppression of small virtuality splitting in the
region of small or moderate virtuality, while we use ordinary fixed order
perturbation theory in the large virtuality region where ``parton
shower'' physics is not relevant.

Another possibility would be to modify ${\cal I}[{\rm shower}]$ by
replacing ${\cal M}$ by a new function ${\cal M}'$ and by replacing
${\cal P}$ by a new function ${\cal P}'$. The new functions could be
anything we like provided that they have the same small $\bar q^2 \to 0$
limits as the original functions. They could, for example, be the
splitting functions in a ``standard'' showering scheme based on virtuality
as the evolution variable. Then, in the order $\alpha_s^2$ graphs, ${\cal
M}$ would be replaced by ${\cal M} - {\cal M}'$ and ${\cal P}$ would be
replaced by ${\cal P} - {\cal P}'$. Thus the shower functions  ${\cal M}'$
and ${\cal P}'$ would act as collinear subtraction terms in the order
$\alpha_s^2$ graphs. This is similar to what is done for initial state
collinear singularities in \cite{FrixioneWebberI,FrixioneWebberII}. The
treatment discussed above with a virtuality cutoff $\lambda_V < \infty$
is a special case of this more general possibility. We do not pursue the
general ${\cal M}'$,
${\cal P}'$ possibility here.

\section{Primary splittings for a whole graph}

\begin{figure}
\includegraphics[width = 8 cm]{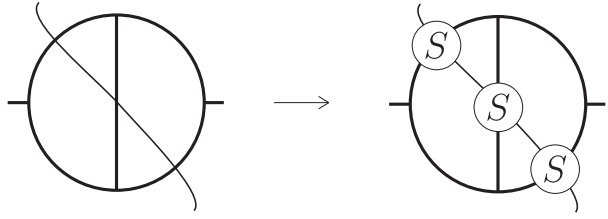}
\medskip
\caption{A cut Born graph. We depict here only the topology of the
graph, leaving unspecified which lines are quarks and which are
gluons. On the right, we show the graph after replacing the cut
propagators by ${\cal I}[{\rm shower}]$, which is represented by a
circle containing an $S$. }
\label{fig:born2}
\end{figure}

\begin{figure}
\includegraphics[width = 10 cm]{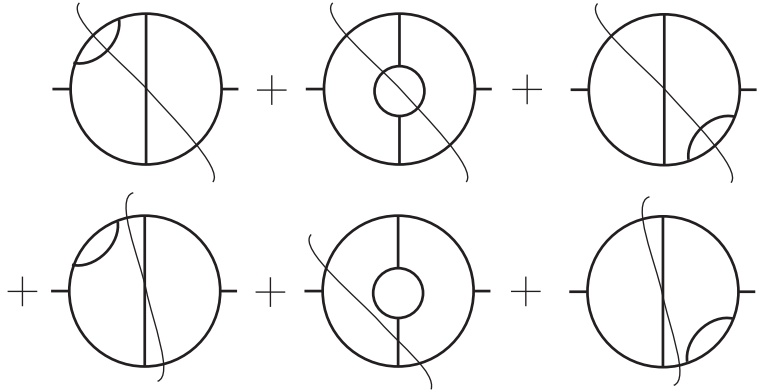}
\medskip
\caption{Six order $\alpha_s^2$ cut Feynman graphs. (For a graph with
a self-energy insertion with one adjacent propagator cut, we understand
the complete virtual correction.) These graphs are the order $
\alpha_s^2$ term in the expansion of the graph in
Fig.~\ref{fig:born2}.}
\label{fig:nlo2a}
\end{figure}

Let us see how this algorithm works in a particular example, taking
$\lambda_V = \infty$ for simplicity. Begin with a cut Born ({i.e.}
order $\alpha_s^1$) graph with the topology shown on the left in
Fig.~\ref{fig:born2}.  The primary splitting algorithm is to replace
$\cal{I}[{\rm Born}]$ by $\cal{I}[{\rm shower}]$ for each of the three
parton lines crossing the cut. This is illustrated in the right hand side
of Fig.~\ref{fig:born2}.

When this expression is expanded in powers of $\alpha_s$, it
gives the Born graph back along with the six order $\alpha_s^2$
graphs depicted in Fig.~\ref{fig:nlo2a}. There are further
contributions that are of order $\alpha_s^3$ and higher. Since we get
the graphs in Fig.~\ref{fig:nlo2a} from the expansion of the right hand
diagram in Fig.~\ref{fig:born2}, we delete these contributions from the
order $\alpha_s^2$ graphs.

\section{Primary splittings for another graph}

\begin{figure}[h]
\includegraphics[width = 10 cm]{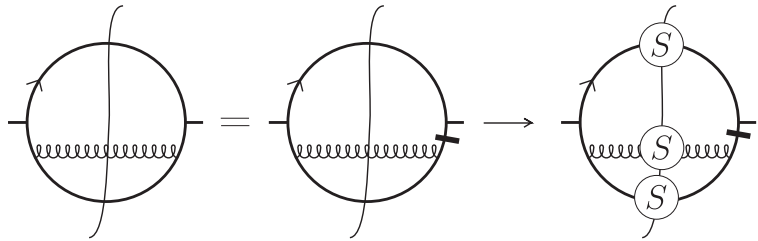}
\medskip
\caption{A cut Born graph and its replacement according to the level 1
showering algorithm.}
\label{fig:born1}
\end{figure}

There is one more topology for a cut Born graph that contributes to
three-jet cross sections. This topology is exemplified by the
graph shown as the left hand diagram in Fig.~\ref{fig:born1}. Again, we
keep this analysis as simple as possible by treating the case that the
virtuality cut is $\lambda_V = \infty$.

To treat this graph, we make use of a technique that helps keep two-jet
physics from affecting the three-jet cross section. Let $k_1$ be the
momentum carried by the top propagator in Fig.~\ref{fig:born1} and
let $k_2$ be the momentum carried by the bottom propagator before it
splits into two propagators. We multiply the graph by 
\begin{equation}
1 = F_1(k_1,k_2) + F_2(k_1,k_2),
\end{equation}
where
\begin{equation}
 F_j(k_1,k_2) = { (k_j^2)^2 \over (k_1^2)^2 + (k_2^2)^2}.
\end{equation}
We separate the two terms and treat them separately, considering the
factor $F_j$ as part of the graph. We can indicate which
term is which by drawing a line through the propagator that carries
momentum $k_j$ when we have multiplied by $F_j$. The line indicates that
the corresponding propagator is blocked from going on-shell because of
the factor $(k_j^2)^2$. For the graph under consideration, only the $F_2$
term contributes, as indicated by the equality between the left and
middle graphs of  Fig~\ref{fig:born1}.

Now we adopt as the splitting algorithm for this case the replacement of
each cut propagator by ${\cal I}[{\rm shower}]$, as indicated by the right
hand graph in Fig~\ref{fig:born1}. It is significant that we do this
only after inserting the factor $F_2$.

The role of the factor $F_j$ is easy to understand on an intuitive basis.
It is possible for the graph on the left in Fig.~\ref{fig:born1} to
generate a two-jet event in which the gluon is nearly parallel to the
antiquark at the bottom of the graph.  When each parton is
allowed to split, one could get a high virtuality splitting from the quark
at the top of the graph, creating a three-jet event. Of course, we want to
generate three-jet events, but our calculation is based on generating
three-jet events directly from the Born graphs. The function $F_j$
enforces the requirement that the high virtuality splitting be the one in
the Born graph.

\begin{figure}[ht]
\includegraphics[width = 12 cm]{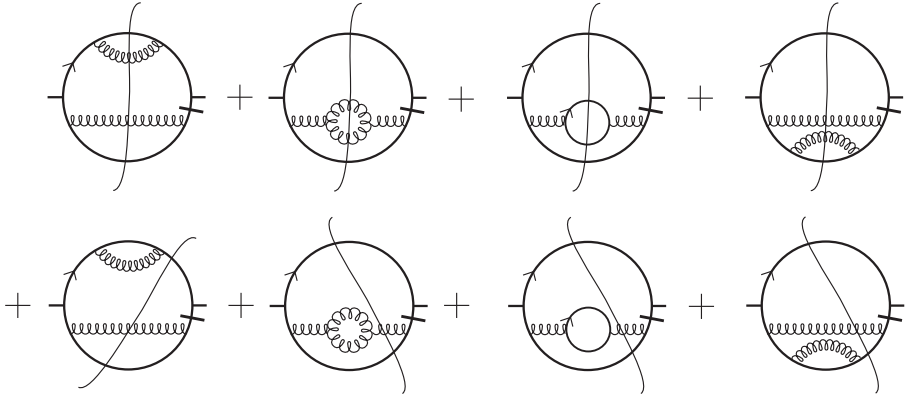}
\medskip
\caption{Eight order $\alpha_s^2$ cut Feynman graphs.}
\label{fig:nlo1a}
\end{figure}

Consider now the eight order $\alpha_s^2$ cut Feynman graphs illustrated
in Fig.~\ref{fig:nlo1a}, where we have inserted factors of $1 = F_1 +
F_2$. For all except the first graph shown, only the $F_2$ term
survives, as indicated. For the first graph, there is a corresponding
graph with a factor $F_1$, which is not shown and would be grouped with
a different set of graphs.

Now notice that when the right hand graph of Fig.~\ref{fig:born1} is
expanded in powers of $\alpha_s$, it reproduces the Born graph with
which we started plus the graphs of Fig.~\ref{fig:nlo1a} plus
corrections that are of order $\alpha_s^3$ or higher. Thus we take the
splitting algorithm for this case to be that the graphs of 
Fig.~\ref{fig:nlo1a} are deleted from the calculation.

\section{The next-to-leading order graphs}

We have seen how to insert a one level parton splitting into the Born
graphs. We have also seen that certain of the order $\alpha_s^2$ cut
graphs are thereby generated, so that these graphs are to be deleted
from the calculation. In the case that we have chosen $\lambda_V <
\infty$, then the graphs that would have been deleted are retained, but
with a virtuality cut $\bar q^2 > \lambda_V \vec q^{\,2}$ on the
appropriate cut self-energy subdiagram. Other order $\alpha_s^2$ cut
graphs are not generated from the Born graphs and need to be added to the
Born contributions.

\section{Secondary splittings}

We have replaced each cut propagator in the Born diagrams
with one step of a parton shower and have called the parton
splittings generated in this way primary splittings. Now we would
like to let these partons split further with secondary
splittings. For each of
the parton lines entering the final state, we replace
$\cal{I}$ for that line by a simplified version of the splitting
probability. For example, if the parton is a quark, we use
\begin{eqnarray}
{\cal I}[{\rm shower\mbox{-}0}]&=&
\int\! {d\vec q\over  2|\vec q\,|}\ {\rm Tr}\Biggl\{ 
\int_0^\infty {d\bar q^2 \over \bar q^2}
\int_0^1\! dx
\int_{-\pi}^\pi\! {d\phi\over 2\pi}\
{ \alpha_s \over 2\pi}\,{\cal S}_{\!g/q}(\bar q^2,x)\,\rlap{/}q\,
\,R(\bar q^2,x,\phi)
\nonumber\\
&&\times
\exp\left(-\int_{\bar q^2}^\infty{ d\bar l^2 \over \bar l^2}
\int_0^1\! dz\
{ \alpha_s \over 2\pi}\,{\cal S}_{\!g/q}(\bar l^2,z)
\right)
\Biggr\},
\label{simpleshower}
\end{eqnarray}
where the Sudakov exponent ${\cal S}_{g/q}$ is a suitable simplified
version of ${\cal P}_{\!g/q}$. Here the Dirac structure, $\rlap{/}q$, is
kept exactly the same as it was in $\cal{I}[{\rm Born}]$ for this line.
We thus take $q^0 = |\vec q|$ in $\rlap{/}q$. Furthermore, we take $q^0 =
|\vec q|$ in the rest of the graph. When this expression is expanded in
powers of $\alpha_s$, it gives the original graph back along with some
order $\alpha_s^3$ and yet higher order corrections.

In the construction so far, we have used a coupling $\alpha_s$ evaluated
at a fixed scale, $\mu$. As a refinement, we follow standard
practice by using for the showering in Eq.~(\ref{simpleshower}) a coupling
evaluated at a running scale. We take the scale of $\alpha_s$ in the main
line of the equation to be $[x (1-x)\bar q^2 S/s_0]^{1/2}$ and in the
exponent to be $[z (1-z)\bar l^2 S/s_0]^{1/2}$. Recall that we use a
numerical integration program that employs dimensionless momenta as
integration variables. The factor $x (1-x)\bar q^2$ is the dimensionless
squared transverse momentum that is associated with the splitting. A
factor $S/s$ would convert to the physical momenta. We use instead
$S/s_0$, where $\sqrt{s_0} = \sum |\vec k_j|$ is the sum of the absolute
values of the dimensionless momenta in the perturbative diagram before
splittings.  Of course, when we introduce a running coupling inside
an integral as in Eq.~(\ref{shower}), we create an ambiguity about how to
evaluate $\alpha_s(k)$ when $k \lesssim \Lambda_{\rm QCD}$. We resolve
the ambiguity by freezing $\alpha_s(k)$, so that it ceases to grow
with decreasing $k$ as soon as it reaches the value $\alpha_s = 1$.

The function ${\cal S}_{\!g/q}(\bar q^2,x)$ should be fairly simple, its
integral in the Sudakov exponent should be finite, and it should approach
the Altarelli-Parisi function $\tilde P_{g/q}(x)$ as $\bar q^2 \to 0$. 
The reader may choose his or her favorite function. We have tried the
following choice, which satisfies these criteria, for the purposes of this
paper,
\begin{equation}
{\cal S}_{\!g/q}(\bar q^2,x) = 
\tilde P_{g/q}(x)\
\theta(\bar q^2 <  x (1-x) \kappa^2).
\label{sdef}
\end{equation}
Here $\kappa^2$ serves as a cutoff parameter for the splitting.
The theta function may be regarded as a simplified version of the factors
$4x(1-x)/[\bar q^2/\vec q^{\,2} + 4  x (1-x)]$ that occur in the
functions ${\cal P}$ that represent the virtual self-energy diagrams and
appear in the Sudakov exponent of the primary splittings. (See the
Appendix for the functions ${\cal P}$.) We examine what to
choose for $\kappa$ below.

We can now move on to generate parton splittings within the previous
parton splittings. We do this indefinitely except that, when $\bar q^2$
for a splitting is smaller than some predetermined cutoff, we regard
the splitting as having gone too far. We then replace the two daughters
by the mother, canceling the splitting. The mother parton remains
in the final state and no further splitting of it is attempted.

Since the mother parton in the current splitting was created in
a previous splitting, one can choose $\kappa^2$ so as to insure
angular ordering between the two splittings.  For a nearly collinear
splitting with parameters $\{\bar q^2,x,|\vec q|\}$ of a mother parton
with momentum $\vec q$, the angle $\theta$ between the two daughter
partons is given approximately by 
\begin{equation}
\theta^2 = { \bar q^2 \over x(1-x) \vec q^{\,2}}.
\end{equation}
Suppose that the previous splitting had parameters $\{\bar p^2,y,|\vec
q|/y\}$, with $y$ being the momentum fraction of the daughter parton that
becomes the mother parton for the second splitting. Then the angle
between the two partons in the first splitting can be estimated as 
\begin{equation}
\theta_0^2 \approx \min\left[ 
{ y\,\bar p^2\over (1-y) \vec q^{\,2}},1
\right].
\end{equation}
The first expression here uses the small angle approximation, while the
second provides a rough upper bound for the case that the angle is not
small. Thus if we take 
\begin{equation}
\kappa^2 = \min\left[ 
{ y\,\bar p^2\over (1-y) }, \vec q^{\,2}, \kappa_{\rm mother}^2
\right],
\label{kappasq}
\end{equation}
the probability for the second splitting is zero when the
second angle is greater than the first. Here $\kappa_{\rm mother}$
is the $\kappa$ parameter that was used for generating the splitting
that gave the mother parton in the current splitting.

We have not yet specified what the splitting scale $\kappa^2$ should be
for the first secondary splitting, the splitting of one of the partons
that was generated in a primary splitting. We should be a bit careful
because the function $R$ used in the proof in Secs.~\ref{sec:selfenergy}
and \ref{sec:onelevel} now includes the measurement function convoluted
with the functions describing the secondary splittings. In the notation
of the preceding paragraph, $R(\bar p^2,y,\phi)$ should approach $R_0$
when $\bar p^2 \to 0$. This will happen if $\kappa^2 \to 0$ (so that the
secondary jets become narrow) when $\bar p^2 \to 0$. For this reason, we
take
\begin{equation}
\kappa^2 = \min\left[ 
{ y\,\bar p^2\over (1-y) }, \vec q^{\,2}, c_\kappa\, \bar p^2
\right],
\label{kappasqstart}
\end{equation}
where $c_\kappa$ is a constant, taken to be 4 in the numerical example
studied in Sec.~\ref{sec:numerical}.

\section{Singularities of the next-to-leading order graphs}

In the next-to-leading order calculation without showers, the order
$\alpha_s^2$ cut graphs contain divergences arising from soft and
collinear parton configurations. These divergences cancel between
different cuts of a given graph, that is between real and virtual parton
contributions. 

In the Coulomb gauge, the collinear divergences arise from
self-energy graphs. When we incorporate parton showers the collinear
divergences disappear. If we use Eq.~(\ref{theorem}), we simply eliminate
the self-energy graphs associated with final state particles, and with
them we eliminate the collinear divergences. With the small modification
in Eq.~(\ref{lambdamodification}), we retain part of the real splitting
self-energy graph in which the virtuality $\bar q^2$ is bigger than
$\lambda_V\vec q^{\,2}$. This still eliminates the collinear divergence.
As mentioned in Sec.~\ref{sec:onelevel}, we could adopt
suitably behaved alternative functions ${\cal M}'$ and ${\cal P}'$ to
replace ${\cal M}$ and ${\cal P}$ in the splitting from a Born level
graph. Then the order $\alpha_s^2$ real and virtual self-energy graphs
would still be present, but ${\cal M}$ would be replaced by ${\cal M} -
{\cal M}'$ in the real graphs and ${\cal P}$ would be replaced by ${\cal
P} - {\cal P}'$ in the virtual graphs. Thus the shower functions  ${\cal
M}'$ and ${\cal P}'$ would act as collinear subtraction terms in the real
and virtual order $\alpha_s^2$ graphs. In this case also, the collinear
divergences are eliminated separately from the real and virtual graphs.

This leaves only the soft divergences in the order $\alpha_s^2$ graphs.
The soft gluon singularities are an essential part of QCD. Their
structure is quite different from that of the collinear singularities.
Their treatment will be the subject of another paper \cite{softshower}.
Once the soft singularities are accounted for, one can let each parton
from an order $\alpha_s^{2}$ graph create a shower. Meanwhile, the
treatment given here, without showers from the order $\alpha_s^2$ graphs,
yields a divergence-free calculation because the soft gluon divergences
cancel between different cuts of a given graph. We will exhibit some
results from this calculation in the following section.

\section{A numerical test}
\label{sec:numerical}

In this section, we provide a numerical test of the algorithm presented
here. We have computed the thrust distribution $d\sigma/dt$ for thrust $t
= 0.86$, in the middle of the three-jet region. We compute the thrust
distribution using the algorithm described in this paper. That is, for
self-energy graphs in the limited virtuality region, we replace ``${\it
Born}\times[ 1 + {\it real} - |{\it virtual}|]$'' by ``${\it Born}\times
{\it real}\,\exp(-|{\it virtual}|)$.'' Secondary showering from the Born
graphs is also included. On the other hand, in this test there are no
showers from the order $\alpha_s^{B+1}$ graphs.

The idea is to compare the thrust distribution thus calculated to the
same distribution calculated with a straightforward NLO computation with
no showers. According to the analysis presented above, the difference
between the two calculations should be of order $\alpha_s^{B+2}$. We
divide the difference by the NLO result, forming
\begin{equation}
R = \frac{\mbox{\rm (NLO-shower)} - {\rm NLO}}{{\rm NLO}}.
\label{Rdef}
\end{equation}
The ratio $R$ should have a perturbative expansion that begins at order
$\alpha_s^2$. We can test this by plotting the ratio against 
$\alpha_s^2$. We expect to see a curve that approximates a straight line
through zero for small $\alpha_s^2$. For comparison, we exhibit also the
ratio with the $\alpha_s^{B+1}$ corrections omitted,
\begin{equation}
R_{\rm LO} = \frac{\mbox{\rm (LO-shower)} - {\rm NLO}}{{\rm NLO}}.
\label{RLOdef}
\end{equation}
The ratio $R_{\rm LO}$ should have a perturbative expansion that
begins at order $\alpha_s^1$. Thus we expect to see a curve proportional
to the square root function $[{\alpha_s^2}]^{1/2}$ for small $\alpha_s^2$.
(The size of the coefficient of $[{\alpha_s^2}]^{1/2}$ in $R_{\rm LO}$
does not have any great significance, since it is quite sensitive to the
choice of renormalization scale.)  A comparison of the two curves, $R$ and
$R_{\rm LO}$, shows whether the $\alpha_s^{B+1}$ corrections, which are
lacking in $R_{\rm LO}$, are doing their job.

\begin{figure}
\includegraphics[width = 8 cm]{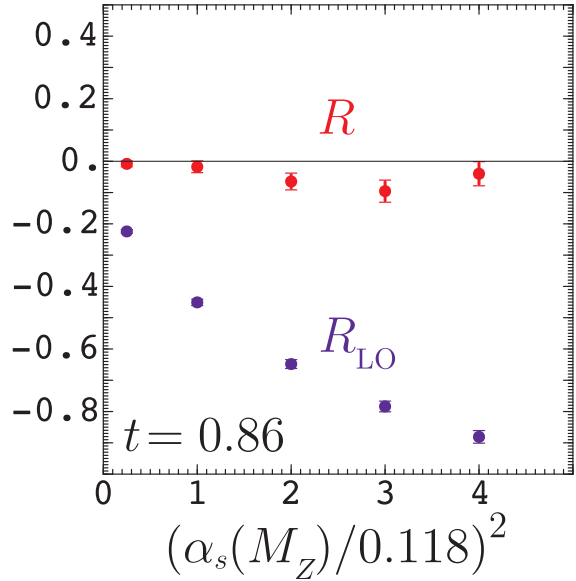}
\medskip
\caption{Comparison of the NLO calculation with showers partially added as
described in this paper to a pure NLO calculation
using \protect\cite{beowulfcode}.  We plot the ratio $R$ defined in
Eq.~(\ref{Rdef}) for the thrust distribution at thrust equal 0.86. Also
shown is the ratio  $R_{\rm LO}$, defined in Eq.~(\ref{RLOdef}), in which
the order $\alpha_s^{B+1}$ correction terms are omitted from the
calculation. The c.m.\ energy is $\sqrt S = M_Z$ and the renormalization
scale is chosen to be $\mu = \sqrt S/6$. These ratios are calculated for
$\alpha_s(M_Z)^2 = \{0.25,1,2,3,4\}\times (0.118)^2$ and plotted versus
$\alpha_s(M_Z)^2/(0.118)^2$.}
\label{fig:test}
\end{figure}

The results of this test are shown in Fig.~\ref{fig:test} for $\sqrt S =
M_Z$ and for a range of $\alpha_s(M_Z)$ from $(1/2)\times 0.118$ to
$2\times 0.118$, where 0.118 represents something close to the physical
value for $\alpha_s(M_Z)$. We choose the renormalization scale to be $\mu
= \sqrt S/6$. One should note that $2\times 0.118$ amounts to
quite a large $\alpha_s$ in this calculation since $\alpha_s(M_Z) =
0.236$ gives $\alpha_s(\mu) = 0.586$.

We see the expected shape of the $R_{\rm LO}$ curve. We also see that the
$R$ curve lies closer to zero, indicating that the $\alpha_s^{B+1}$
corrections are acting in the right direction. The more exacting test is
to examine whether the $R$ curve approaches a straight line through the
origin as $\alpha_s^2 \to 0$. Within the errors, it does. However, the
slope of the straight line is quite small. That is presumably an accident
of the choice of parameters in the program. It remains for the future to
examine the effects of various parameter choices. 

\begin{figure}
\includegraphics[width = 8 cm]{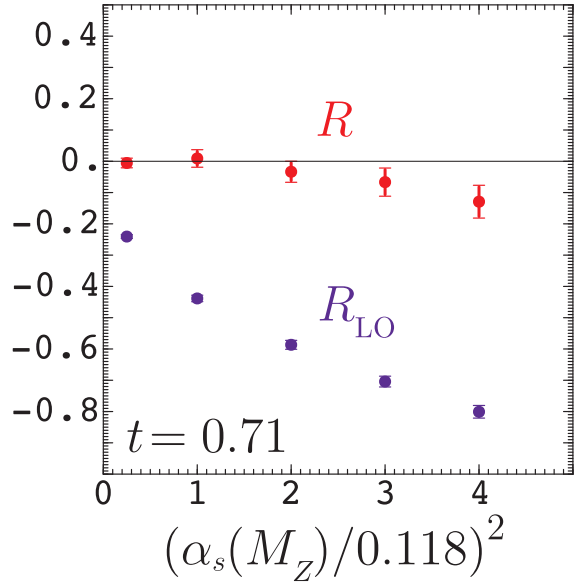}
\medskip
\caption{ Comparison of the NLO calculation with showers partially added to
a pure NLO calculation for $t = 0.71$. The notation is as in
Fig.~\ref{fig:test}.}
\label{fig:test71}
\end{figure}

In Fig.~\ref{fig:test71}, we show the same comparison, but this time for
$t = 0.71$. This is near the value $t = 2/3$ that marks the far end of the
three-jet region, with the three partons in what is sometimes called
the Mercedes configuration. In this case, the $R$ curve again approaches a
straight line through the origin as $\alpha_s^2 \to 0$, again with quite
a small slope.

\begin{figure}
\includegraphics[width = 8 cm]{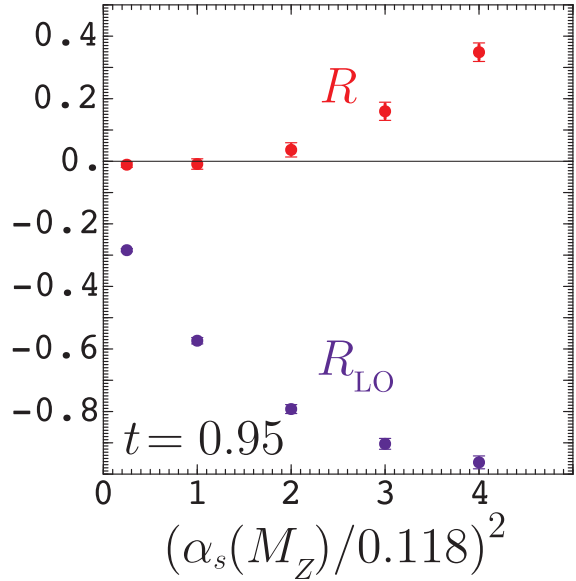}
\medskip
\caption{ Comparison of the NLO calculation with showers partially added to
a pure NLO calculation for $t = 0.95$. The notation is as in
Fig.~\ref{fig:test}.}
\label{fig:test95}
\end{figure}

In Fig.~\ref{fig:test95}, we show the same comparison, but this time for
$t = 0.95$. This is near the two-jet limit at $t = 1$. For $\alpha_s =
0.118$, one would normally not use a calculation that did not include a
summation of logs of $1-t$ for $t$ this close to 1, since $\log(0.05)^2
\approx 9$. Thus we would not recommend using the code discussed in this
paper for a comparison to data this near to the two-jet limit.
Nevertheless, we can still test for the absence of an $\alpha_s^{1}$ term
in $R$. Looking at the graph, we see that, within the errors, there is no
evidence for a nonzero $\alpha_s^{1}$ term in $R$. We expect an
$\alpha_s^{2}$ term, but the coefficient of $\alpha_s^{2}$ appears to be
quite small. On the other hand, it appears that some of the higher order
terms are quite substantial.

\section{Conclusions}

We have presented a method for adding parton showers to a
next-to-leading order calculation in QCD. Specifically, we treat the
final state collinear singularities, using the example of $e^+ + e^- \to
3\ {\rm jets}$.  Our intention is that a program like ours would be used
in conjunction with a standard Monte Carlo event generator
\cite{Pythia, Herwig, Ariadne}.  The complete calculation would then act as an
event generator (with positive and negative weights for the events) in which
the final states consist of hadrons generated from parton showers. Recall
that we defined primary parton splittings to be those of the partons
emerging from an order $\alpha_s^{B}$ graph. There are also secondary
splittings: the further splittings of the daughters from the primary
splittings as well as the splittings of partons emerging from an order 
$\alpha_s^{B+1}$ graph and of their daughters. It is the primary
splittings that need to be matched to the NLO calculation. The secondary
splittings affect an infrared safe observable only at order
$\alpha_s^{B+2}$. Thus the partonic events from our program after the
primary splittings could be handed to a Monte Carlo event generator,
which would perform the secondary splittings and the hadronization.
Alternatively, our program can perform the secondary splittings, leaving
hadronization for the Monte Carlo event generators. For the purpose of
implementing such a scheme, it is significant that the weight functions
do $\it not$ have singularities when two partons are collinear or one is
soft. This is in contrast to pure NLO calculations, which depend on
cancellations between different events when an infrared safe observable
is calculated.

The method presented here is incomplete in that we leave for a companion
publication \cite{softshower} the treatment of the soft gluon
singularities of QCD. The code  described in this paper is available at
\cite{beowulfcode}. A treatment of soft gluon effects, to be described in
\cite{softshower}, is included in the code. Parton showers are included,
but the requisite data structures and color flow information that would
be necessary to provide an interface with the showering and hadronization
models of standard event generator Monte Carlo programs are not yet
available.

\acknowledgements

The authors are pleased to thank Steve Mrenna for advice on how to
implement parton splitting in a finite amount of computer time. We also
thank John Collins and Torbj\"orn Sj\"ostrand for helpful advice and
criticisms. This work was supported in part by the U.S.~Department of
Energy, by the European Union under contract HPRN-CT-2000-00149, and by
the British Particle Physics and Astronomy Research Council.

\appendix*

\section{Virtual splitting functions}
\label{sec:calP}

In this appendix, we examine the functions ${\cal P}_{\!a/b}(\bar q^2,x)$
that specify the virtual parton self-energy graphs in the Coulomb
gauge.

For quark splitting, the function that we use is 
\begin{eqnarray}
{\cal P}_{\!g/q}(\bar q^2,x) &=&
{C_F\over 2 D_\mu}\,
\left\{ 12 x(1-x) - 1\right\}
\nonumber\\
&&+ 
{C_F \over 2 D_\mu^2}\, 
{\bar q^2 \over \mu_{\rm mod}^2}
\left\{
20 x(1-x) - 3
+ 
\log(\mu^2 / \mu_{\rm mod}^2)
[12 x(1-x) - 1]
\right\}
\nonumber\\
&&+{4 C_F \over D_x}\, x(1-x)\left\{5  - 14 x(1-x) \right\}
\nonumber\\
&&+ {16 C_F \over D_x^2}\,x(1-x)
\left\{1 -  6 x(1-x) + 8 [x(1-x)]^2\right\}
\nonumber\\
&&+
{C_F\over 2 D_1}\,(2x-1)
\nonumber\\
&&
- {8 C_F \over D_1 D_x} (2x-1) x(1-x)
\nonumber\\
&&
- {16 C_F \over D_1 D_x^2} (2x-1)  x(1-x) 
\left\{1 - 2 x(1-x)\right\},
\label{calPgq}
\end{eqnarray}
where $\mu$ is the \MSbar\ renormalization scale,
\begin{equation}
\mu_{\rm mod}^2 = \min(\mu^2,|\vec q\,|^2 ),
\label{mdef}
\end{equation}
and
\begin{eqnarray}
D_x &=& \bar q^2/|\vec q\,|^2 + 4 x (1-x) ,
\nonumber\\
D_\mu &=& \bar q^2/\mu_{\rm mod}^2 + 1 ,
\nonumber\\
D_1 &=& \bar q^2/|\vec q\,|^2 + 1 .
\label{Ddef}
\end{eqnarray}
The function ${\cal P}_{\!g/q}(\bar q^2,x)$ has been modified from the
function ${\cal P}_{\!q}$ given in Ref.~\cite{KSCoulomb}. We have replaced
the two terms
\begin{equation}
-{ C_F \over 2}\,{ 1 \over e^{-3}\bar q^2/\mu^2 + 1}
+ C_F { 6 x(1-x) \over e^{-5/3}\bar q^2/\mu^2 + 1}
\label{badtermsq}
\end{equation}
by the first two terms in Eq.~(\ref{calPgq}). The contributions of these
terms to the integral
\begin{equation}
\int_{\bar q^2}^\infty{ d\bar l^2 \over \bar l^2}\,
{\cal P}_{\!g/q}(\bar l^2,x)
\label{pertativeexample}
\end{equation}
are the same in the limit of small $\bar q^2$ up to terms that vanish as
$\bar q^2 \to 0$ (and are non-singular as functions of $x$). Thus the two
expressions are equivalent when inserted into the perturbative formula
(\ref{pert}). 

The terms that we have modified contain the effects of \MSbar\
renormalization. In the original form, they cut off the integral
(\ref{pertativeexample}) at $\bar l^2$ of order $\mu^2$.  That is a
sensible way to express perturbative renormalization. However, in our
present application the integral 
\begin{equation}
\int_{\bar q^2}^\infty{ d\bar l^2 \over \bar l^2}\,
\int_0^1 dx\,
{\cal P}_{\!g/q}(\bar l^2,x)
\label{sudakovexample}
\end{equation}
appears in the Sudakov exponent for shower generation. In the case that
$|\vec q\,|^2 \ll \mu^2$, the former terms in Eq.~(\ref{badtermsq})
would have caused the Sudakov exponential to deviate from 1 for $\bar q^2$
much greater than $|\vec q\,|^2$. The effect of this deviation on the
calculated cross section is of order $\alpha_s^2 \times R_0$ and thus is,
in principle, negligible. However this effect appears to us to be quite
unphysical. The revised form of ${\cal P}_{\!g/q}$ is preferable in that
it confines the contribution of the renormalization dependent terms to
$\bar q^2$ of order $|\vec q\,|^2$ or smaller.

One may also note that the integral (\ref{sudakovexample}) is positive
for all $\bar q^2$ and increases as $\bar q^2$ decreases. This was not the
case with the form of ${\cal P}_{\!g/q}$ before the replacement. 

For gluon splitting to a quark and antiquark, we use
\begin{eqnarray}
{\cal P}_{\!q/g}(\bar q^2,x) &=&
{1\over 2 D_\mu}\,
\left\{1 - 2 x(1-x) \right\}
\label{calPqg} 
\\
&&+ 
{1 \over 2 D_\mu^2}\, 
{\bar q^2 \over \mu_{\rm mod}^2}
\left\{
2 - {\textstyle{16\over 3}}\,x(1-x)
+ 
\log(\mu^2/\mu_{\rm mod}^2)
[1 - 2 x(1-x)]
\right\}.
\nonumber
\end{eqnarray}
For gluon splitting to two gluons, we use
\begin{eqnarray}
{\cal P}_{\!g/g}(\bar q^2,x) &=&
{C_A\over D_\mu}\,
\left\{ 3 x(1-x) - 1\right\}
\nonumber\\
&&+ 
{C_A \over D_\mu^2}\, 
{\bar q^2 \over \mu_{\rm mod}^2}
\left\{
7\, x(1-x) - 2
+ 
\log(\mu^2 / \mu_{\rm mod}^2)
[ 3 x(1-x) - 1]
\right\}
\nonumber\\
&&+{12 C_A \over D_x}\, x(1-x)\left\{1  - 2 x(1-x) \right\}
\nonumber\\
&&+ {16 C_A \over D_x^2}\,x(1-x)
\left\{1 -  4 x(1-x) + 4 [x(1-x)]^2\right\}.
\label{calPgg}
\end{eqnarray}
Here the definitions (\ref{mdef}) and (\ref{Ddef}) apply again. For 
${\cal P}_{\!q/g}$, we have replaced 
\begin{equation}
{ 1 \over 2}\,{ 1 \over e^{-2}\bar q^2/\mu^2 + 1}
- {  x(1-x) \over e^{-8/3}\bar q^2/\mu^2 + 1}
\end{equation}
given in Ref.~\cite{KSCoulomb} by the terms in Eq.~(\ref{calPqg}). For 
${\cal P}_{\!g/g}$, we have replaced 
\begin{equation}
-  C_A \,{ 1 \over e^{-2}\bar q^2/\mu^2 + 1}
+ C_A {  x(1-x) \over e^{-5/3}\bar q^2/\mu^2 + 1}
+ 2 C_A{  x(1-x) \over e^{-8/3}\bar q^2/\mu^2 + 1}
\end{equation}
given in Ref.~\cite{KSCoulomb} by the first two terms in
Eq.~(\ref{calPgg}). The reasoning behind this modification is the same as
in the case of quark splitting.

The Sudakov exponent for gluon splitting is proportional to
\begin{equation}
\int_{\bar q^2}^\infty{ d\bar l^2 \over \bar l^2}\,
\int_0^1 dx\,
\left[
{\cal P}_{\!g/g}(\bar l^2,x)
+ N_F \,{\cal P}_{\!q/g}(\bar l^2,x)
\right].
\label{sudakovexampleg}
\end{equation}
This quantity is positive for all $\bar q^2$ and increases as $\bar q^2$
decreases provided that $N_F > 3 C_A/2$. Thus this form will work fine
for the main cases of phenomenological interest, $C_A = 3$ with 5 or 6
quark flavors.  Further modifications would be necessary for other cases.
We have not pursued this topic further because we expect that the $\cal
P$ functions will be different anyway in future versions of the
algorithms presented in this paper.

\end{document}